\documentclass[amsmath,amssymb,twocolumn]{revtex4-1}
\usepackage{graphicx}
\usepackage{hyperref}
\usepackage{url}
\begin{document}
%%%%%%%%%%%%%%%%%%%%%%%%%%  Title  %%%%%%%%%%%%%%%%%%%%%%%%%%
\title{Real-time phasefront detector for heterodyne interferometers}
%%%%%%%%%%%%%%%%%%%%%%%%%%  Authors  %%%%%%%%%%%%%%%%%%%%%%%%%%

\author{Felipe Guzm\'an Cervantes}
\email[Corresponding author:]{felipe.guzman@aei.mpg.de}
\author{Gerhard Heinzel}
\author{Antonio F. Garc\'ia Mar\'in}
\author{Vinzenz Wand}
\author{Frank Steier}
\author{Karsten Danzmann}
\affiliation{Max-Planck-Institut f\"ur Gravitationsphysik (Albert-Einstein-Institut Hannover), and Institut f\"ur Gravitationsphysik, Leibniz Universit\"at Hannover, Callinstra\ss e 38, 30167 Hannover, Germany}
\author{Oliver Jennrich} 
\affiliation{ESA-ESTEC, Keplerlaan 1, Postbus 299, 2200 AG Noordwijk, The Netherlands}
%%%%%%%%%%%%%%%%%%%%%%%%%%  Abstract  %%%%%%%%%%%%%%%%%%%%%%%%%%
\begin{abstract}
We present a real-time differential phasefront detector sensitive to better than 3\,mrad rms, which corresponds to a precision of about 500\,pm. This detector performs a spatially resolving measurement of the phasefront of a heterodyne interferometer, with heterodyne frequencies up to approximately 10\,kHz. This instrument was developed as part of the research for the LISA Technology Package (LTP) interferometer, and will assist in the manufacture of its flight model. Due to the advantages this instrument offers, it also has general applications in optical metrology.\bigskip
\end{abstract}
\maketitle
%%%%%%%%%%%%%%%%%%%%%%%%%%  Body  %%%%%%%%%%%%%%%%%%%%%%%%%%
\section{Introduction}
\label{intro}
Many applications in optical metrology require precision measurements and characterization of laser beam wavefronts, as well as an accurate mode-matching of laser beams. To this end, it is usual to perform various adjustments that are both complex and time-consuming, based on repeated measurements of beam parameters\,\cite{siegman}. Alternative methods are Shack-Hartmann sensors, which measure the shape of a single wavefront with an accuracy of typically \texttt{$\lambda/100$}\,\cite{haso}, and phase-shifting interferometry\,\cite{lai,millerd} (PSI), which reaches typically \texttt{$\lambda/1000$}\,\cite{kaiser}.\\
Similar to PSI, the relative phase is found by a mathematical algorithm\,\cite{surrel,hibino} which is applied on the intensities sampled for \texttt{$n\geq4$} instantaneous operating points with equidistant phase increments. This happens in parallel at each pixel of the spatially resolving photodetector (CCD camera). The advantage of our method over conventional PSI is that the phase-shift, normally implemented by a piezo-electric transducer (PZT), is intrinsically contained in the time-dependent sinusoid obtained from the interference of two electric fields at different frequencies (optical heterodyning). Thus, non-linearities of PZT-elements, and additional electronics required to keep the interferometer at a specific operation point can be avoided\,\cite{yongqian}.\\
This article describes an apparatus able to measure in real-time the relative spatial structure between two interfering wavefronts in a heterodyne interferometer. This allows a quick and very precise characterization of the mode-mismatch between two beams in real-time, making possible an online adjustment of the optical components according to the mode-mismatch displayed. By using beams with known shape, this method can also be used to analyze surfaces and optical components (in transmission and reflection), reaching a sensitivity better than \texttt{$\lambda/2000$}, which corresponds in our case to a precision of about 500\,pm at \texttt{$\lambda=1064\,\mathrm{nm}$} (Nd:YAG laser).\\
Optical heterodyne interferometry\,\cite{wyant,fengzhao} is a useful technique to measure distance variations with sub-wavelength precision and large dynamic range. This concept is applied, for example, in the LISA Technology Package (LTP)\,\cite{ghh1} which utilizes a set of heterodyne Mach-Zehnder interferometers to measure relative changes in the separation of two drag-free test masses with a noise level better than 10\,\texttt{$\mathrm{pm}/\sqrt{\mathrm{Hz}}$} in the frequency range of 3\,mHz to 30\,mHz. It is well-known that heterodyne interferometers are susceptible to various noise sources\,\cite{zhao,vwa} which corrupt the phase measurement. One important noise source is the wavefront imperfections of the interfering beams in combination with beam jitter\,\cite{ghh3,dave}.
This effect is of particular importance if quadrant photodiodes (QPD) are used, which are often employed to obtain alignment signals from the interferometer. The error term induced by the spatial inhomogeneity of the wavefronts can be minimized if the interfering beams have identical shape. The device we present here allows one to match the shape of the wavefronts with a simple procedure, permitting an online adjustment of the optical components.

\section{Theoretical Background}
\label{theo}

\subsection{Interference pattern and heterodyne interferometry}
\label{het-int}

The electric field \texttt{$\textbf{E}_j\left(\textbf{r},t \right)$} of a linearly polarized light beam can be described as
\begin{equation}
\textbf{E}_j\left(\textbf{r},t \right) =E_j\,\textbf{p}_j\, \mathrm{exp}\left\lbrace\,\mathrm{i}\, \left[\, 2\pi\, f_j\, t+\varphi_j+\psi_j\left(\textbf{r} \right)\,\right]\,\right\rbrace,
\label{e-field}
\end{equation}
where \texttt{$j$} is an index to distinguish several beams, the vector \texttt{$\textbf{p}_j$} describes the polarization, \texttt{$E_j$} is the amplitude of the electric field, and \texttt{$\psi_j\left(\textbf{r}\right)$} is the spatial distribution of the phasefront.  The intensity distribution \texttt{$I\left(\textbf{r},t \right)$} oscillates at the heterodyne frequency \texttt{$f_\mathrm{het}=f_\mathrm{2}-f_\mathrm{1}$}, and is proportional to \texttt{$\mid\textbf{E}_{\mathrm{total}}\left(\textbf{r},t \right)\mid^2$}, where \texttt{$\textbf{E}_\mathrm{total}\left(\textbf{r},t \right)= \textbf{E}_\mathrm{1}\left(\textbf{r},t \right)+\textbf{E}_\mathrm{2}\left(\textbf{r},t \right)$} is the interference pattern. Assuming identical polarization vectors \texttt{$\textbf{p}_\mathrm{1}=\textbf{p}_\mathrm{2}$} and also that the differential coherence length of the laser used is much larger than the distance from the source to the recombination point of the interferometer, the heterodyne component can be described as
\begin{equation}
I\left(\textbf{r},t \right)=A(\textbf{r})\,\left\lbrace\,1+C(\textbf{r})\, \mathrm{cos}\left[\,2\pi\, f_\mathrm{het}\, t-\varphi(\textbf{r}) \,\right] \,\right\rbrace,
\label{intensity}
\end{equation}
where \texttt{$A(\textbf{r})$} is a space-dependent factor, \texttt{$C(\textbf{r})$} describes the contrast in terms of the space \textbf{r}, and
\begin{equation}
\varphi(\textbf{r})=\left[ \, \varphi_1+\psi_1\left(\textbf{r} \right)\,\right] -\left[\, \varphi_2+\psi_2\left(\textbf{r} \right)\,\right] 
\end{equation}
gives the spatial dependence of the phase, where \texttt{$\psi\left(\textbf{r} \right)=\psi_1\left(\textbf{r} \right)-\psi_2\left(\textbf{r} \right)$} is the spatially distortion of the phase distribution. Ideally, for identically shaped wavefronts (\texttt{$\psi_1\left(\textbf{r}\right)=\psi_2\left(\textbf{r}\right)$}), the longitudinal phase term \texttt{$\varphi_1-\varphi_2$}, which is the usual interferometric quantity to be measured, contains the length measurement describing the pathlength difference \texttt{$\Delta L$} between the arms of the interferometer:
\begin{equation}
\Delta L= \frac{\lambda}{2\pi}\,\left (\varphi_1-\varphi_2\right),
\label{delta-l}
\end{equation}
where \texttt{$\lambda=c/f$} is the wavelength of the light. This pathlength difference can be strongly influenced by environmental changes that usually (as in LTP) disturb the main length measurement.\\
In order to maximize the contrast of the interferometer, the beams need to be matched. It is therefore very useful to characterize the mismatch between them, and to be able to optimize it in real-time.

\subsection{Spatially resolved phase measurement}
\label{phase-meas}

The relative geometry of two interfering beams can be described by the spatial structure of the functions \texttt{$A(\textbf{r})$}, \texttt{$C(\textbf{r})$}, and \texttt{$\psi\left(\textbf{r}\right)$}. The apparatus described here measures these three functions in the real-time interference pattern using a CCD camera and pixelwise data processing. In order to obtain the phase of a sinusoidal function as in Equation~(\ref{intensity}), several mathematical approaches\,\cite{surrel} can be used which are based on measuring \texttt{$n$} equidistant intensity samples \texttt{$I_k=I(t_k)$}, with \texttt{$t_k=k\,\Delta t$}, where \texttt{$k$} is an integer. The approach chosen for this experiment is a 4-point algorithm\,\cite{freischald}, corresponding to a straightforward Discrete Fourier Transform (DFT) of the signal with \texttt{$n=4$} samples and \texttt{$\Delta t=T/4$}, where \texttt{$T=1/f$} is the period of the signal. For a noise-free signal, these intensities would be given by
\begin{equation}
I_k=I_\mathrm{avg}\,\left[\,1+C\,\mathrm{cos}\left(\varphi+k\,\frac{\pi}{2}\right)\,\right],
\label{ik}
\end{equation}
where \texttt{$I_\mathrm{avg}$} is the average of the sampled intensities.\\
The phase at the pixel \texttt{$\gamma$}, \texttt{$\varphi_\gamma$}, can be calculated from these intensity samples \texttt{$\left( I^{(\gamma)}_0\dots I^{(\gamma)}_3\right)$} with the 4-point algorithm as
\begin{equation}
\varphi_\gamma=\mathrm{arctan}\left(\frac{I^{(\gamma)}_3-I^{(\gamma)}_1}{I^{(\gamma)}_0-I^{(\gamma)}_2}\right).
\label{phase}
\end{equation}
Several other useful quantities can also be obtained from these 4 data points. Using the abbreviations
\begin{eqnarray}
a_\gamma &=& I^{(\gamma)}_0-I^{(\gamma)}_2\label{a}\\
b_\gamma &=& I^{(\gamma)}_3-I^{(\gamma)}_1\label{b}\\
d_\gamma &=& I^{(\gamma)}_0+I^{(\gamma)}_1+I^{(\gamma)}_2+I^{(\gamma)}_3\label{d},
\end{eqnarray}
we get:
\begin{itemize}
\item Contrast at the pixel \texttt{$\gamma$}:
\begin{equation}
C_\gamma=2\cdot\frac{\sqrt{a_\gamma^2 +b_\gamma^2 }}{d_\gamma}
\label{contrast}
\end{equation}
\item Total phase over the CCD surface:
\begin{equation}
\varphi_\mathrm{total}=\mathrm{arctan}\left( \frac{\sum_\gamma b_\gamma }{\sum_\gamma a_\gamma }\right)
\label{total-phase}
\end{equation}
\item Total contrast over the CCD surface:
\begin{equation}
C_\mathrm{total}=2\cdot \frac{\sqrt{\left(\sum_\gamma a_\gamma\right)^2+\left(\sum_\gamma b_\gamma\right)^2}}{\sum_\gamma d_\gamma}
\label{total-contrast}
\end{equation}
\item Average intensity at the pixel \texttt{$\gamma$}:
\begin{equation}
I^{(\gamma)}_\mathrm{avg}=\frac{d_\gamma}{4}
\label{avg}
\end{equation}
\item The maximum and minimum intensity for the set of exposures (\texttt{$I_0\dots I_3$}) can also be determined for diagnostic purposes.
\end{itemize} 
Furthermore, an exposure of the dark fringe can be directly captured by triggering the CCD camera with the appropriate delay \texttt{$\tau_\mathrm{df}$}:\\
\begin{equation}
\tau_\mathrm{df}=\frac{3\pi/2-\varphi_\mathrm{total}}{2\pi}\, T_\mathrm{het}.
\label{dark-fringe}
\end{equation}

\section{Instrument components and setup}
\label{phasemeter}

\begin{figure}
\begin{center}
\includegraphics[width=0.95\columnwidth]{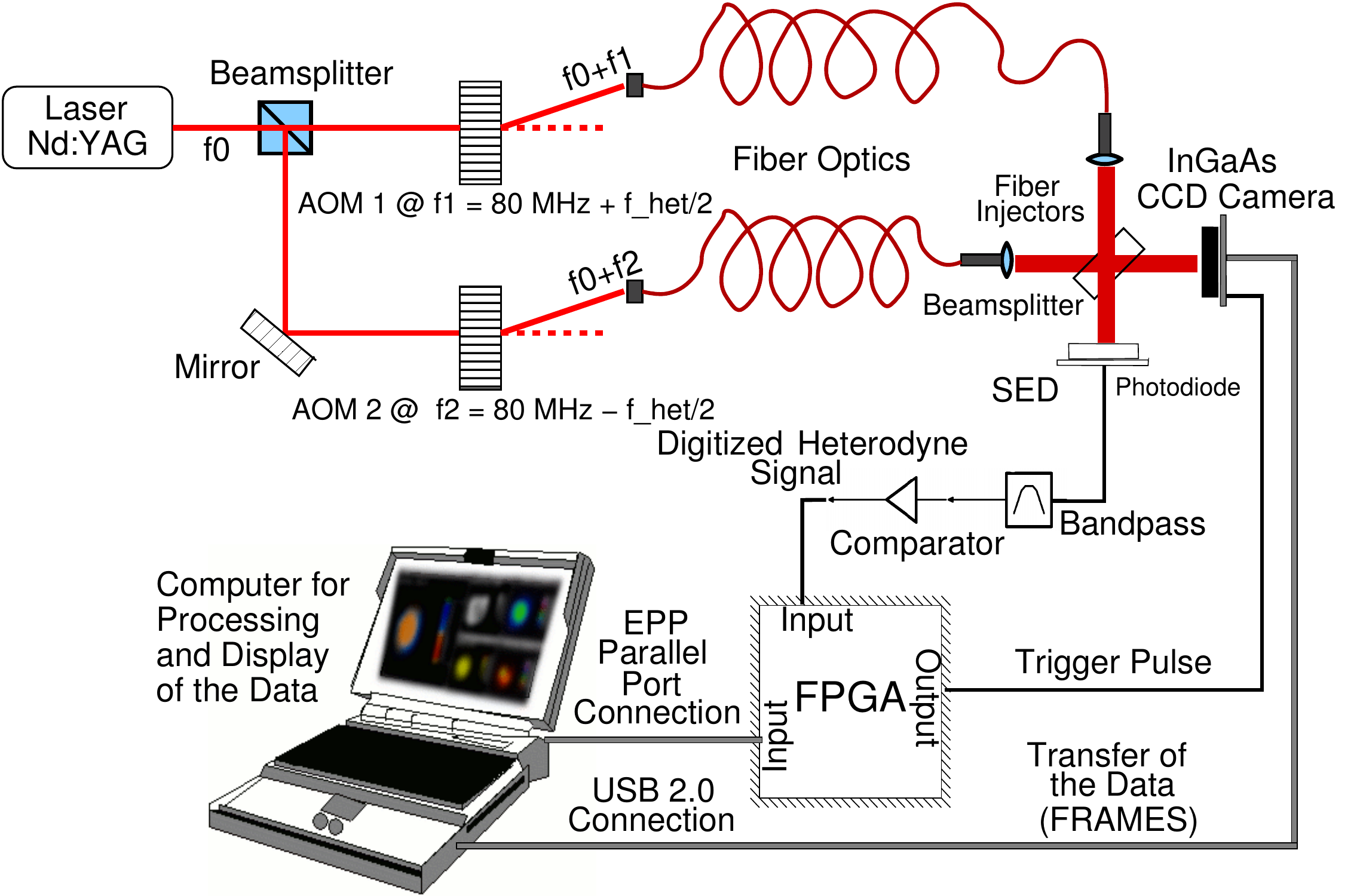}
\caption{\label{setup}Experimental setup used for the phasemeter.}
\end{center}
\vspace{-0.6cm}
\end{figure}

\begin{figure}
\begin{center}
\includegraphics[width=0.95\columnwidth]{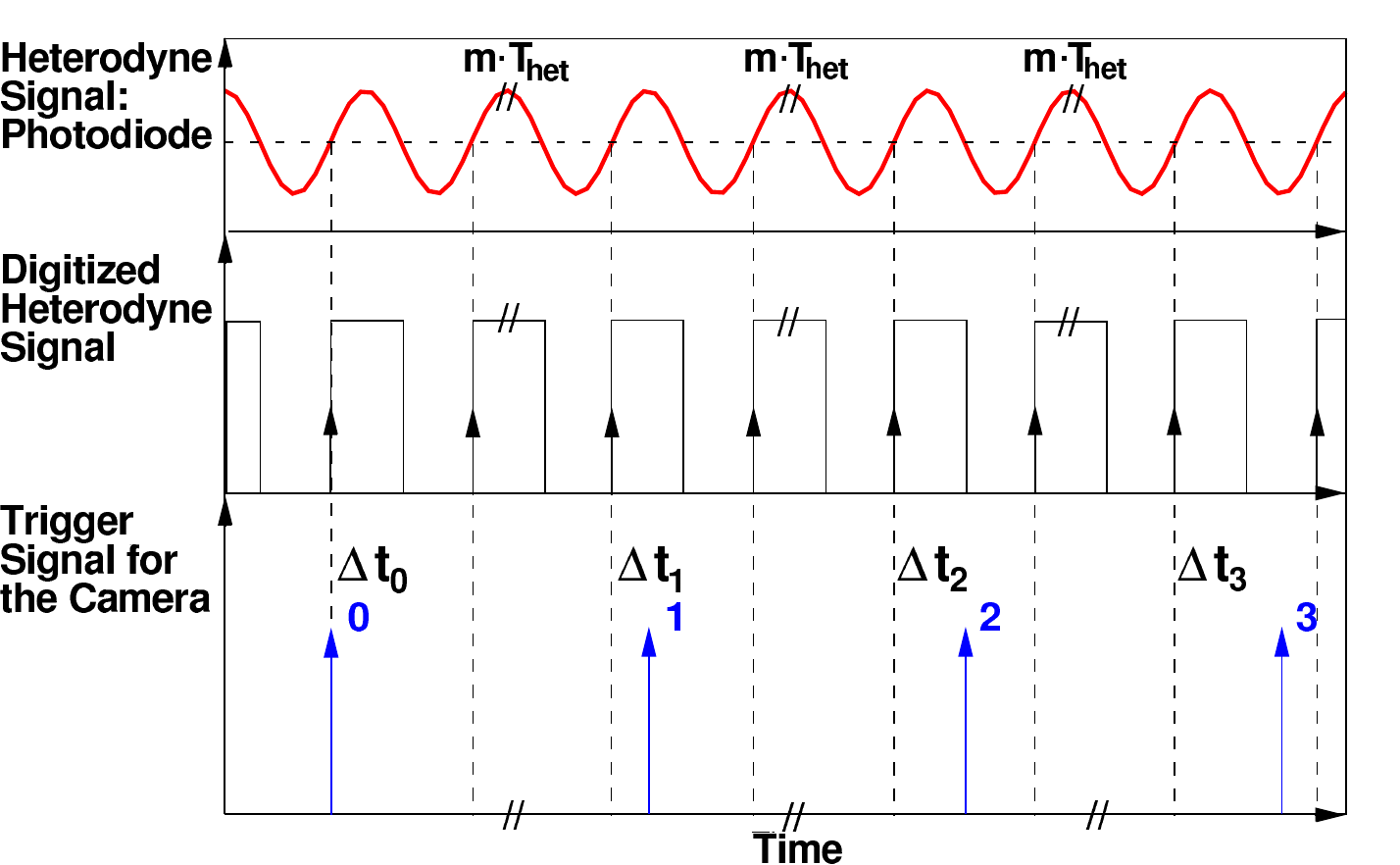}
\caption{\label{trigger}Time diagram of the signals processed to trigger the CCD camera. Note that the sinusoid is measured by the SED and yields a 180$^\circ$ phase shift wrt. to the interference pattern measured by the CCD camera.}
\end{center}
\vspace{-0.6cm}
\end{figure}

The interferometer configuration used in this experiment is a non-polarizing heterodyne Mach-Zehnder interferometer. The light source was a Nd:YAG NPRO (non-planar ring oscillator) laser with a wavelength of 1064\,nm. Two acousto-optic modulators (AOM), driven by slightly frequency shifted RF signals near 80\,MHz, are used to generate two laser beams with a frequency difference of \texttt{$f_\mathrm{het}\approx1623\,$}Hz. Since the beams diffracted by an AOM have a distorted non-Gaussian beam profile, single-mode polarization-maintaining fiber optics are used as mode cleaners. The requirements on the CCD camera are:
\begin{itemize}
\item Simultaneous exposure for every pixel (\lq\lq global shutter\rq\rq\,).
\item The exposure time must be very short compared with the heterodyne period, \texttt{$T_\mathrm{het}\approx1/1623\,\mathrm{Hz}=616\,\mu$}s.
\item The camera must be able to be triggered externally to allow frames to be captured at the required instances of time.
\item The signal for each pixel needs to be proportional to the intensity on that pixel. Saturation effects, such as \lq\lq blooming\rq\rq\,, must be avoided.
\end{itemize}
The CCD camera used is a model XEVA-USB from XenICs\,\cite{xenics} with a 12-bit dynamic range which reaches 30 frames per second (fps) at a resolution of \texttt{$320\times 256$} pixels with a pixel pitch of 30\,\texttt{$\mu$}m. The photosensitive chip is made of InGaAs, which has a quantum efficiency of approximately 80\% for the near infrared (0.9--1.7\,\texttt{$\mu$}m). The exposure time used was 1\,\texttt{$\mu$}s\,(\texttt{$1/616\, T_\mathrm{het}$}). Ideally, the interference pattern should be sampled 4 times within a single heterodyne period of \texttt{$T_\mathrm{het}$}. This would require a sampling period of $\Delta t=154\,\mu$s (approximately 6500\,fps) for the CCD camera which cannot be reached in practice due to the time required to transfer the image. In order to capture the intensity sample \texttt{$I_k$}, an integer number \texttt{$m$} of heterodyne periods \texttt{$T_\mathrm{het}$} is added to \texttt{$\Delta t$}, and the trigger signal is sent to the CCD camera with the delay
\begin{equation}
\Delta t_k= m\,T_{\mathrm{het}}+\tau_k.
\label{delta-t}
\end{equation}
with \texttt{$\tau_k=k\, T_\mathrm{het}/4$}. The heterodyne period \texttt{$T_\mathrm{het}$} is estimated by the timing control electronics of a Field Programmable Gate Array (FPGA) at the beginning of the measurement as an average over 2000 periods, and is then transfered to the PC through the parallel port interface. The experimental setup is outlined in Figure~\ref{setup}.
In practice, the environment is not stable enough to preserve a constant phase relationship over many periods of \texttt{$f_\mathrm{het}$}. Hence, additional circuitry is used to re-synchronize the trigger timing electronics with the actual phase of the heterodyne signal: A single-element photodiode (SED) is located at the second output of the interference beamsplitter where the same interference pattern emerges with a \texttt{$180^\circ$} phase shift. The heterodyne signal measured by the SED is bandpass filtered and digitized by a comparator. When the phasemeter software is ready to capture the frame \texttt{$k$}, a command is sent from the PC to the FPGA through the parallel port, which includes the corresponding delay \texttt{$\tau_k$}. The FPGA detects the rising edge of the digital heterodyne signal and triggers the camera with the controlled delay \texttt{$\tau_k$} (see Figure~\ref{trigger}).
The CCD camera captures the frame and transfers it to the PC through its USB 2.0 interface. After all four frames have been acquired, the phasemeter software computes the physical quantities described in Equations~(\ref{phase}) to~(\ref{avg}), and an additional exposure is captured by triggering the camera with a delay \texttt{$\tau_\mathrm{df}$} given by Equation~(\ref{dark-fringe}), which corresponds to an exposure of the dark fringe.
\section{Results}
\label{results}
\subsection{Real-time operation}
\begin{figure}[!ht]
\begin{center}
\includegraphics[width=\columnwidth]{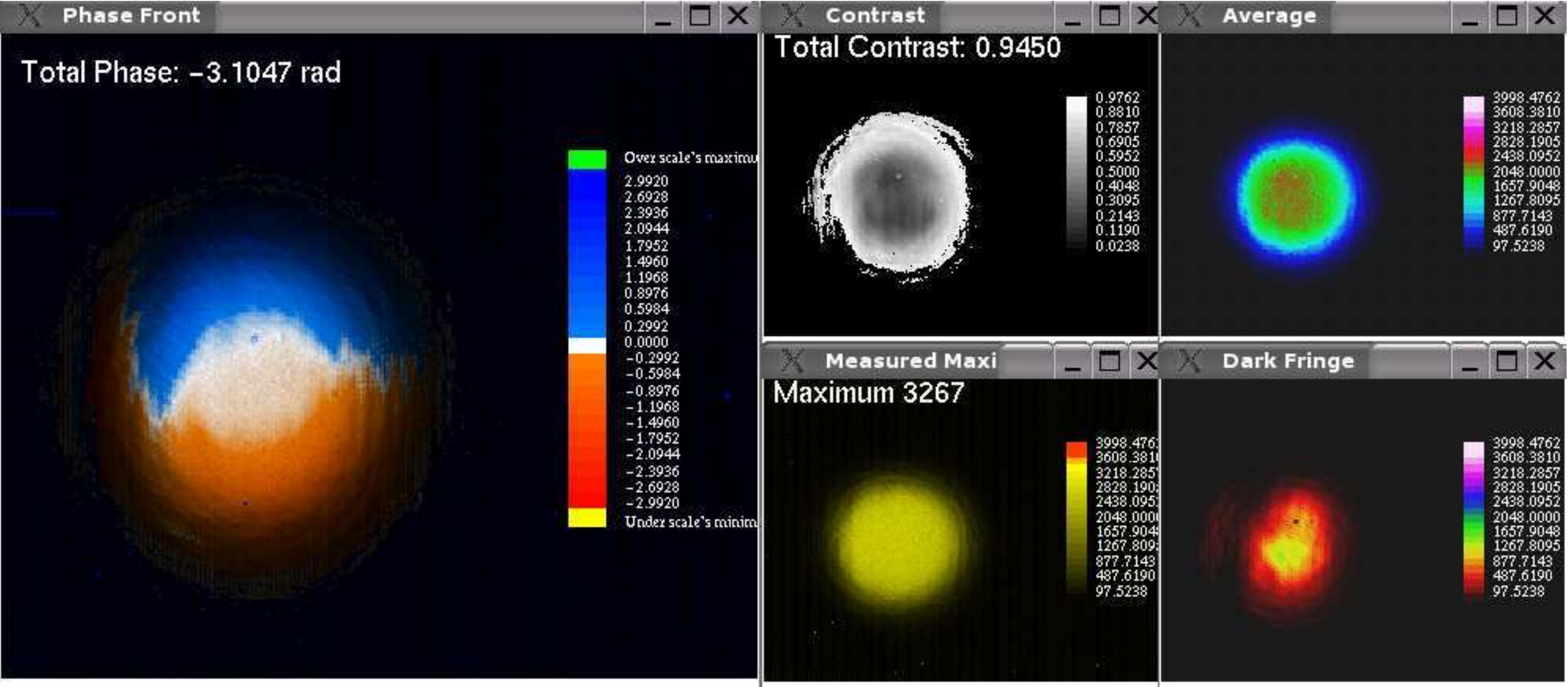}
\caption{\label{gui}Graphical User Interface programmed to display the measured data in real-time.}
\end{center}
\end{figure}

\begin{figure}[!ht]
\begin{center}
\includegraphics[width=0.74\columnwidth]{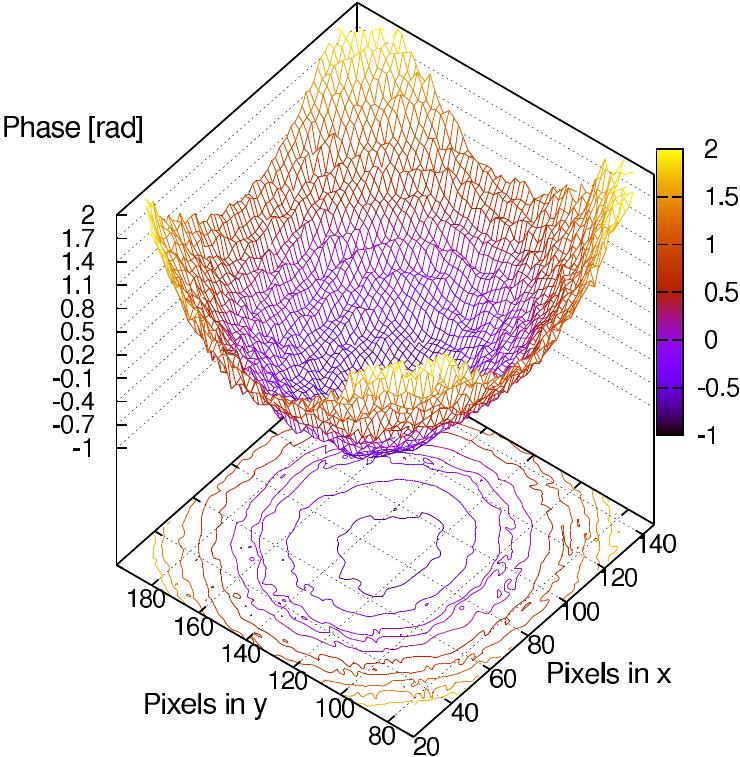}
\caption{\label{phase-3d}Spatial distribution of the phase.}
\end{center}
\end{figure}

\begin{figure}[!ht]
\begin{center}
\includegraphics[width=0.74\columnwidth]{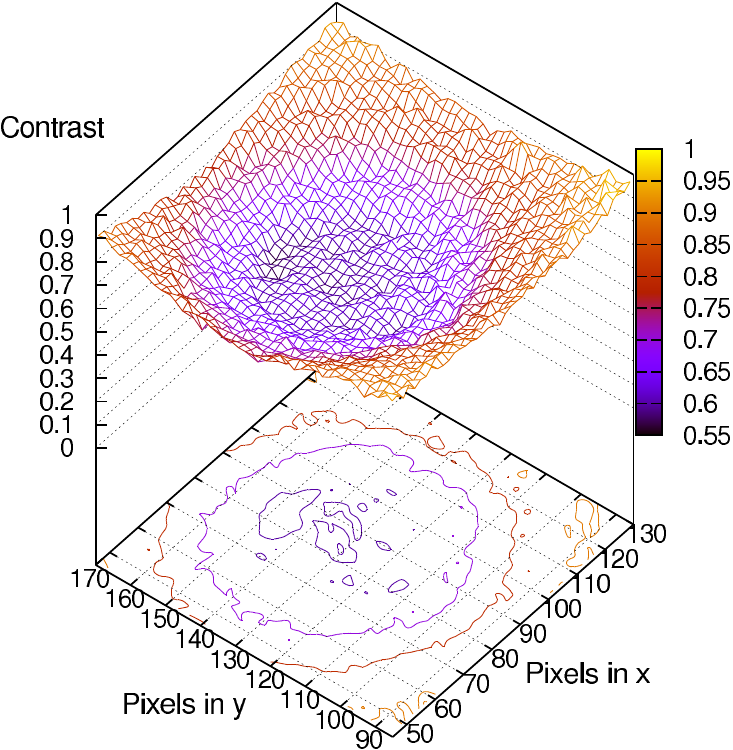}
\caption{\label{con-3d}Spatial distribution of the contrast.}
\end{center}
\end{figure}

\begin{figure}[!ht]
\begin{center}
\includegraphics[width=0.74\columnwidth]{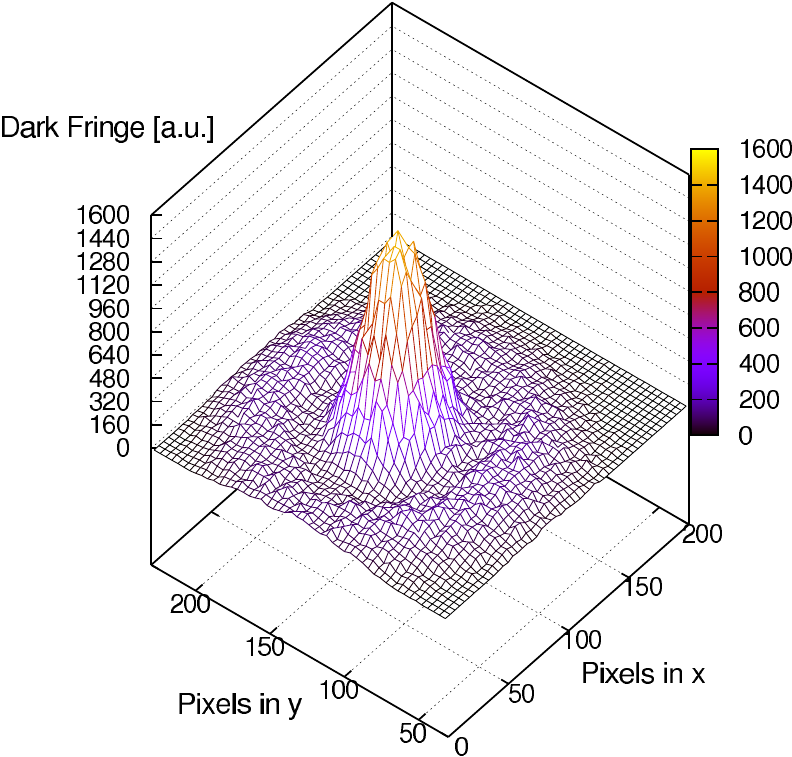}
\caption{\label{df-3d}Exposure of a dark fringe.}
\end{center}
\end{figure}

\begin{figure}[!ht]
\begin{center}
\includegraphics[width=0.74\columnwidth]{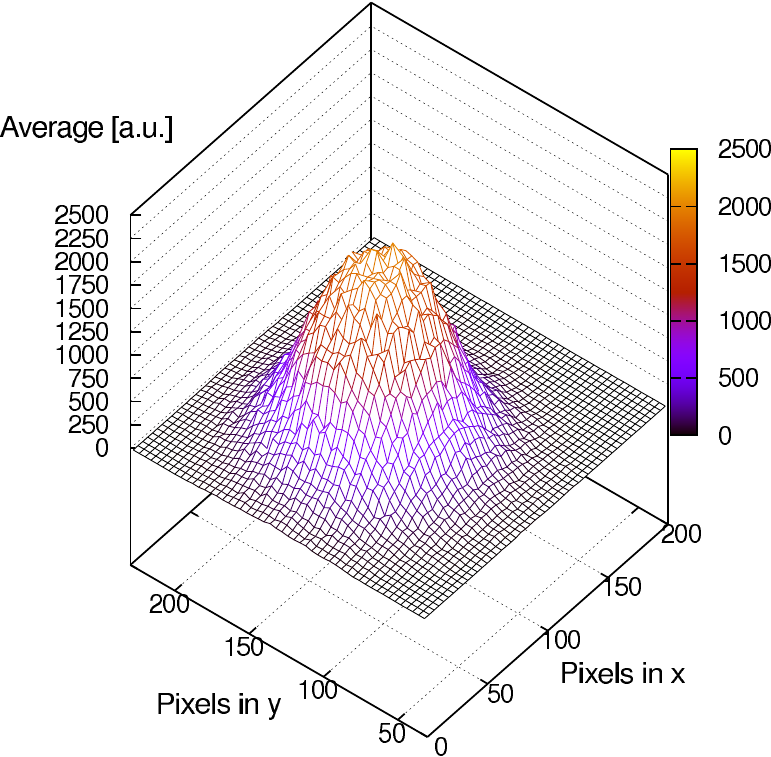}
\caption{\label{avg-3d}Average intensity over four exposures.}
\end{center}
\end{figure}

\begin{figure}[!ht]
\begin{center}
\includegraphics[width=0.74\columnwidth]{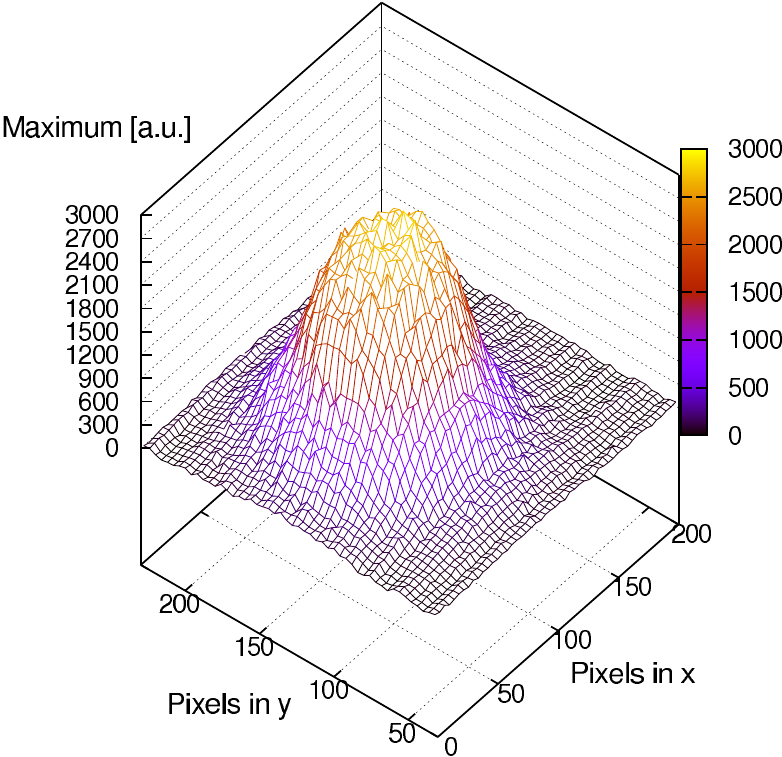}
\caption{\label{max-3d}Maximum intensity over four exposures.}
\end{center}
\end{figure}

A graphical user interface (GUI) was developed to display the measured data in real-time (see Figure~\ref{gui}).
The phasemeter reaches a rate of approximately 5 to 6 data displays per second, and hence enables real-time mode-matching of the beams, as well as an online optical alignment of an interferometer and optimal adjustment of its components. The five different displays in Figure~\ref{gui} are shown separately as 3D representations in Figures \ref{phase-3d} to \ref{max-3d}.
These measurements were conducted on a table-top Mach-Zehnder interferometer, as shown in Figure~\ref{setup}.
\subsection{Performance of the phasemeter}
In order to measure the noise level of the instrument, the camera was illuminated with a spatially homogeneous light source (an array of infrared LED's behind a matt glass window), the intensity of which was modulated sinusoidally at a frequency of approximately 1623\,Hz. This is a very stable test subject, ideal to determine the phase noise level of the instrument. It is expected to obtain a flat phasefront of the amplitude modulation from this measurement, since all pixels capture a similar sinusoidally modulated intensity with the same phase relationship. A series of real-time phasefront measurements were conducted in this configuration, determining the rms phase value at every pixel. Thus, a spatial rms phase variation of 2.96\,mrad was meaured over the CCD area of observation. According to Equation~(\ref{delta-l}), this value corresponds to a rms spatial resolution \texttt{$\Delta L$} of about 500\,pm.\\
As it can be seen in Equation~(\ref{phase}), one noise source of the phase measurement is the fluctuation of the sampled intensities. The rms error of the phase, \texttt{$\Delta\varphi_{\mathrm{rms}}$}, induced by intensity fluctuations \texttt{$\Delta I_{\mathrm{rms}}$} can be estimated from Equation~(\ref{phase}) as:
\begin{eqnarray}
\Delta\varphi_{\mathrm{rms}} & = &\sqrt{\sum_k \left( \frac{\partial\varphi}{\partial I_k}\right)^2}\,\Delta I_{\mathrm{rms}}\\ \nonumber
& = &\sqrt{\frac{2}{\left(I_0-I_2\right)^2+\left(I_1-I_3\right)^2}}\,\Delta I_{\mathrm{rms}}.
\label{phi-error}
\end{eqnarray}
After simplifying Equation~(\ref{phi-error}) by using Equation~(\ref{ik}), we obtain:
\begin{equation}
\Delta\varphi_{\mathrm{rms}}=\frac{\sqrt{2}}{C}\,\frac{\Delta I_{\mathrm{rms}}}{I_\mathrm{avg}}.
\label{phi-error-rms}
\end{equation}
The following three error sources were identified, and their noise contribution to the phase measurement was estimated:
\begin{enumerate}
\item Laser power fluctuations: An Allan deviation of \texttt{$8.6\times10^{-4}$} was measured at an averaging time of 33\,ms, which corresponds to the sampling period of the CCD camera (30\,fps), yielding a phase error of 1.22\,mrad from Equation~(\ref{phi-error-rms}).
\item ADC digital noise of the camera: A rms intensity error of four quantization units was measured by constant and spatially homogeneous illumination of the CCD camera. The pixelwise rms variation and an average over the CCD surface were then computed. This value corresponds to a relative intensity fluctuation of \texttt{$9.76\times10^{-4}$}, which translates (by using Equation~(\ref{phi-error-rms})\,) into a phase error of 1.38\,mrad.
\item Time jitter: There are at least three sources of jitter. Firstly, the synchronization delay of the comparator output with respect to the 10\,MHz clock of the FPGA, which is uniformly distributed between 0 and 100\,ns. Secondly, a similar delay between the FPGA clock and the CCD internal clock, which is also at 10\,MHz but unsynchronized. Thirdly, other jitter effects such as apparent period fluctuations of the signal, due to setup conditional and limited stability, and the non-sychronization between the FPGA and CCD clocks with the clock of the modulation electronics controlling the AOM's driving signal, and therefore the heterodyne frequency generation. This latter effect is reduced (but not totally cancelled) by re-tracking the timing control electronics to the rising edge of the digital heterodyne signal for each exposure. A phase error of 0.99\,mrad was obtained by simulating the first and second effects in software, using two independent random delays uniformly distributed.
\end{enumerate}
Table~\ref{noise-proj} summarizes the noise contributions identified for the phase measurement.
\begin{table}[hb]
\begin{center}
\textbf{ \caption{\label{noise-proj}Main noise sources of the phase measurement.} }
\begin{tabular}{l c}\hline
\textbf{Noise Source} & \textbf{RMS Phase Error} \\ \hline
Laser power fluctuations & 1.22\,mrad \\
ADC digital noise of the camera & 1.38\,mrad \\
Time jitter & 0.99\,mrad \\
\textbf{Total contribution} & \textbf{2.09\,mrad} \\
\textbf{Noise level measured} & \textbf{2.96\,mrad} \\ \hline
\end{tabular}
\end{center}
\end{table}
\begin{figure}[!ht]
\begin{center}
\includegraphics[width=\columnwidth]{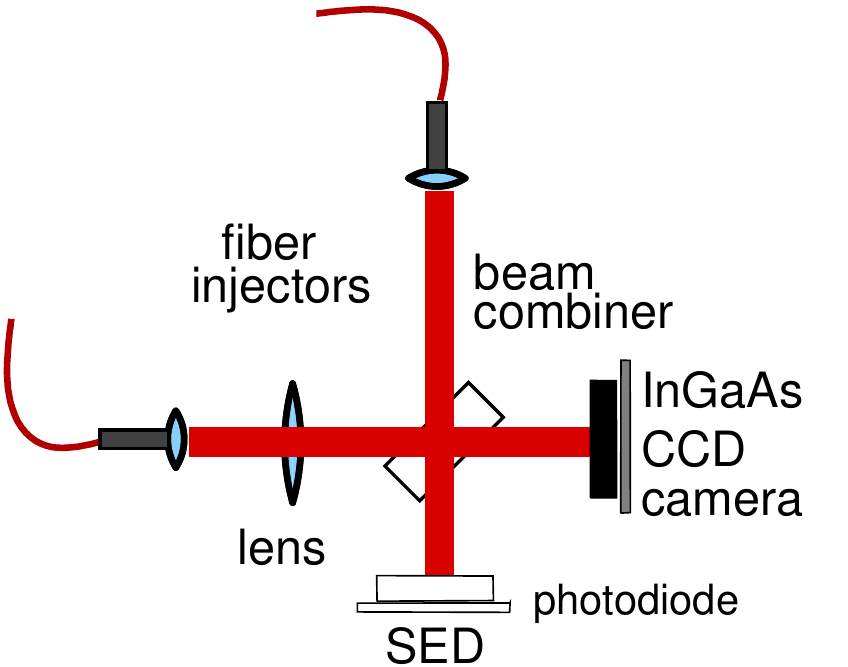}
\caption{\label{setup-lens}Experimental setup with an additional lens in the path of one beam to intentionally change the curvature of its wavefront.}
\end{center}
\end{figure}
\begin{figure*}
\begin{center}
\begin{tabular}{c c}
\includegraphics[width=0.8\columnwidth]{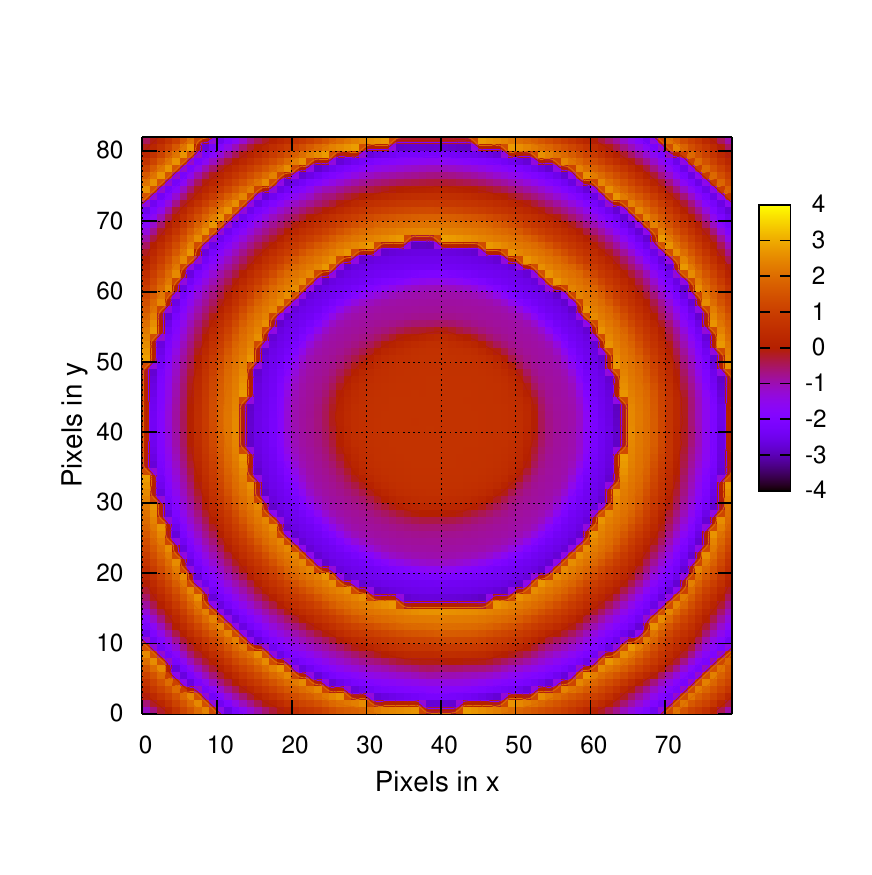} & \includegraphics[width=0.8\columnwidth]{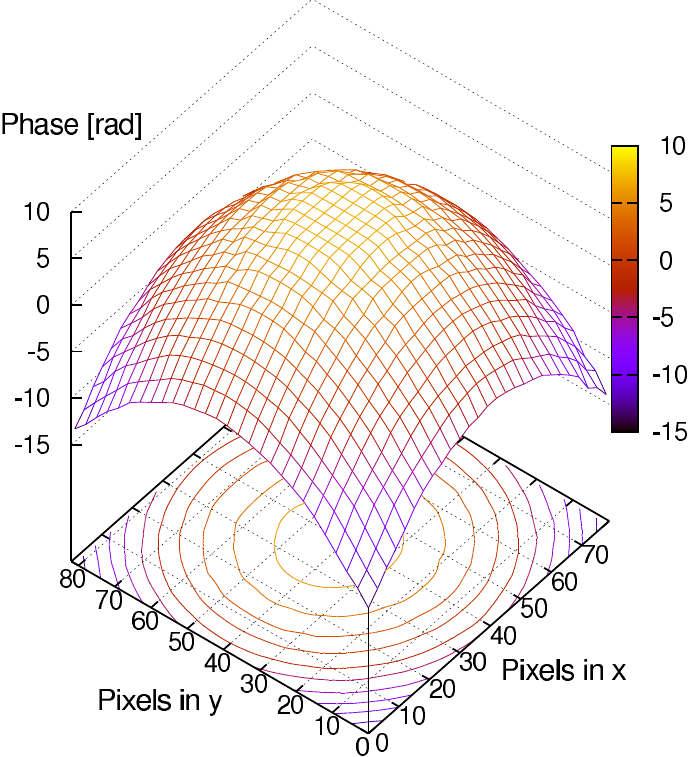}\\
(a) & (b)
\end{tabular}
\end{center}
\caption{\label{rings_down}(a) Phase front measured with a lens \texttt{$f\,=\,+\,500\,$}mm in one arm of the interferometer. The phasefront is clearly wrapped, due to the high curvature of the wavefront being transmitted through the lens wrt. the other one. (b) Phase front obtained by post-processing the data measured in Figure~\ref{rings_down}(a) with a two-dimensional phase unwrapping algorithm.}
\end{figure*}
\begin{figure*}
\begin{center}
\begin{tabular}{c c}
\includegraphics[width=0.8\columnwidth]{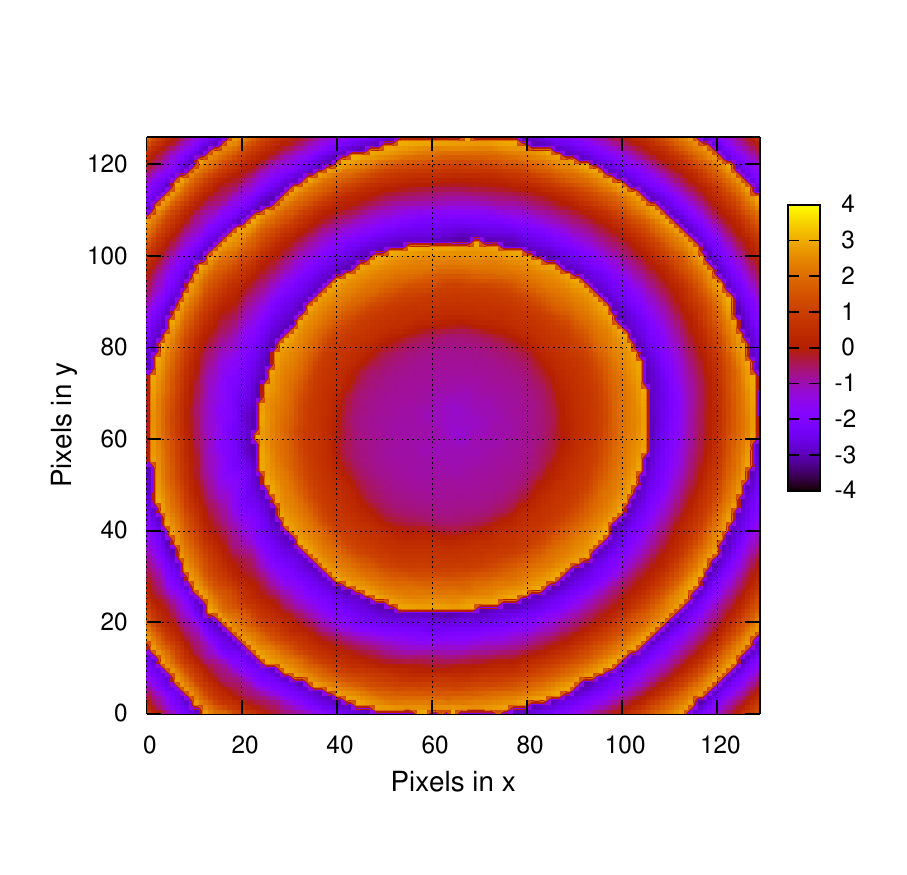} & \includegraphics[width=0.8\columnwidth]{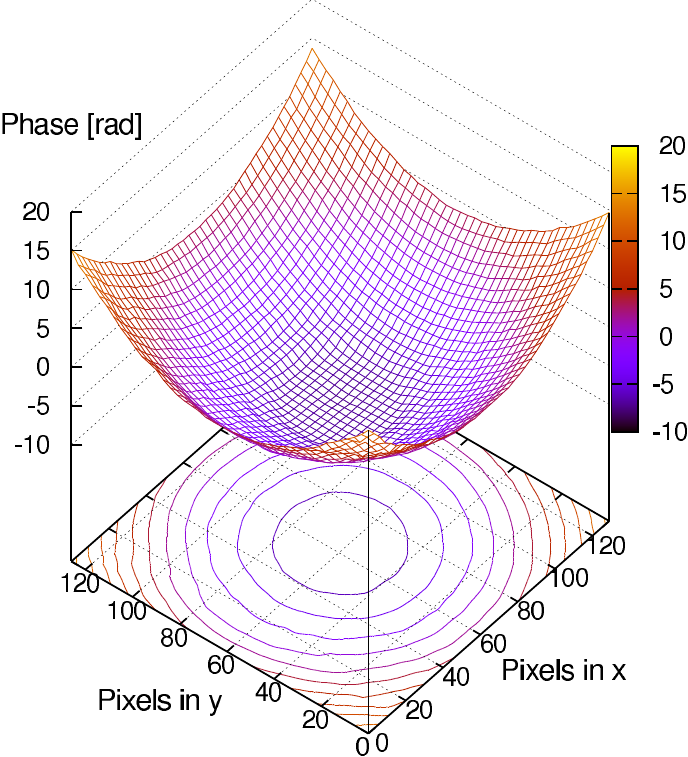}\\
(a) & (b)
\end{tabular}
\caption{\label{rings_up}(a) Phase front measured with a lens \texttt{$f\,=\,-\,500\,$}mm in one arm of the interferometer. (b) Phase front obtained from post-processing the data of Figure~\ref{rings_up}(a) with a 2D phase unwrapping algorithm.}
\end{center}
\end{figure*}
\begin{figure*}
\begin{center}
\begin{tabular}{c c}
\includegraphics[width=0.8\columnwidth]{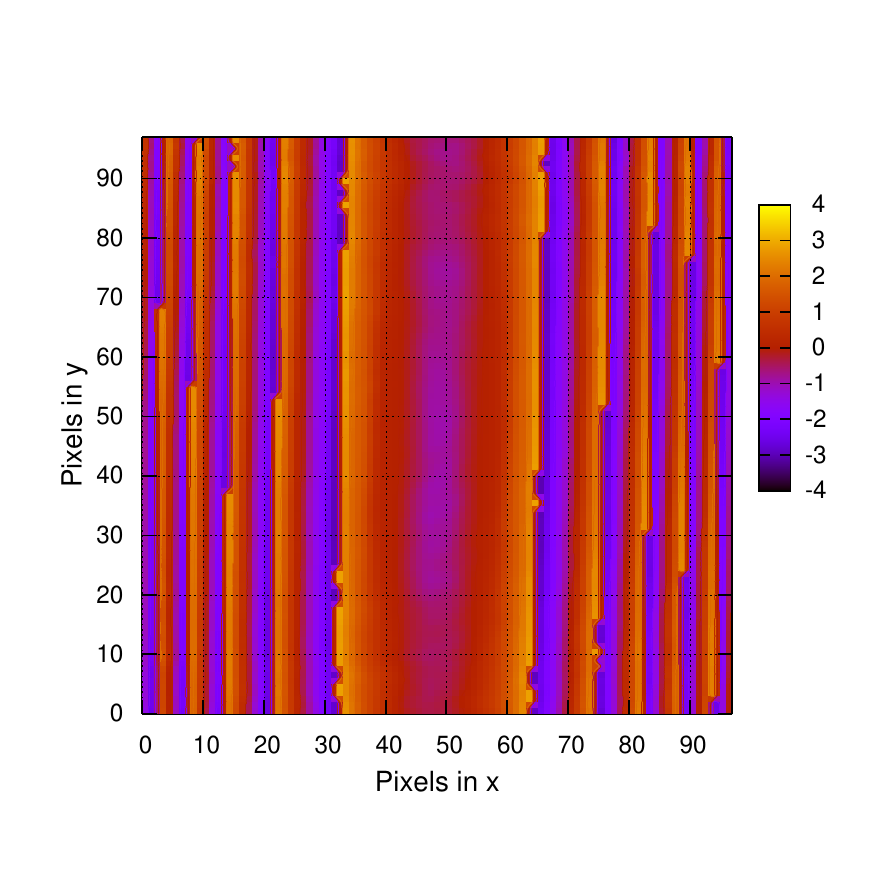} & \includegraphics[width=0.8\columnwidth]{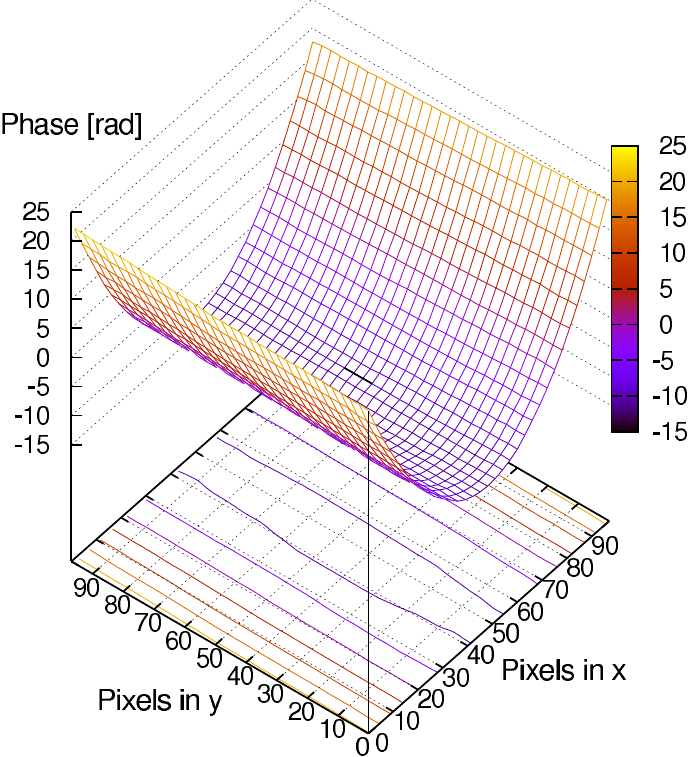}\\
(a) & (b)
\end{tabular}
\caption{\label{cylindric}(a) Phase front measured with a cylindrical lens \texttt{$f\,=\,+\,80\,$}mm in one arm of the interferometer. (b) Phase front obtained from post-processing the data of Figure~\ref{cylindric}(a) with a 2D phase unwrapping algorithm.}                                                     \end{center}
\end{figure*}
\subsection{Measurements}
\begin{figure}
\begin{center}
\includegraphics[width=0.79\columnwidth]{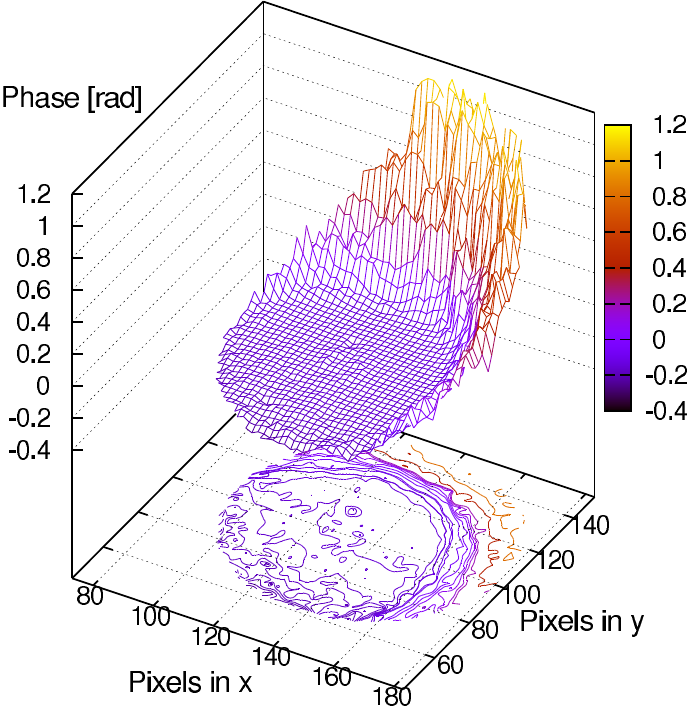}
\caption{\label{ob-phase}Phase front measured at the engineering model of the optical bench for LTP.}
\end{center}
\end{figure}

\begin{figure}
\begin{center}
\includegraphics[width=0.81\columnwidth]{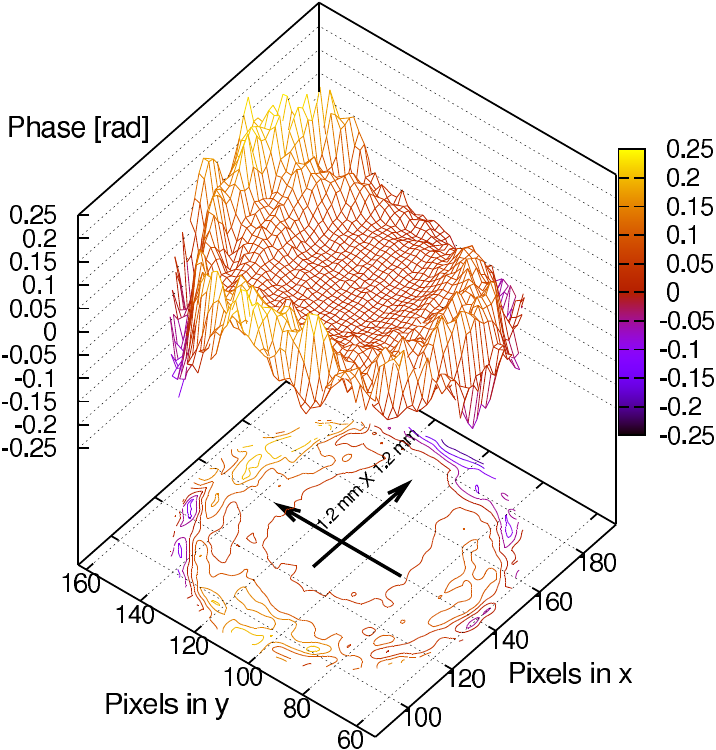}
\caption{\label{flat}Adjusted phasefront measured on a table-top Mach-Zehnder interferometer.}
\end{center}
\end{figure}

In order to test the functionality of the phasemeter, a series of measurements were conducted at the table-top interferometer shown in Figure~\ref{setup}.
The aim of these measurements was to intentionally change the curvature of one of the interfering wavefronts and to use the phasemeter to read out the spatial distribution of the phase. Since the two interfering wavefronts are similar, a lens was introduced into the path of one beam, between the corresponding fiber injector and the beam combiner (see Figure~\ref{setup-lens}).
The results of these measurements are shown in Figures~\ref{rings_down} to~\ref{cylindric}. 
Three different type of lenses were used: \texttt{$f\,=\,+\,500\,$}mm (Figure~\ref{rings_down}), \texttt{$f\,=\,-\,500\,$}mm (Figure~\ref{rings_up}), and a cylindrical lens with \texttt{$f\,=\,+\,80\,$}mm (Figure~\ref{cylindric}). The curvature of the wavefront considerably increases (due to the lens) such that this covered several wavelengths in the area of observation. The resulting phasefront measured is wrapped as can be recognized by the phase rings in Figures~\ref{rings_down}(a) and~\ref{rings_up}(a), and the stripes in Figure~\ref{cylindric}(a). A two-dimensional phase unwrapping algorithm\,\cite{ghiglia1994,flynn,ghiglia1998} was developed and was used for post-processing these data. The result of this post-processing is shown in Figures~\ref{rings_down}(b),~\ref{rings_up}(b), and~\ref{cylindric}(b). It can be seen, by comparing Figures~\ref{rings_down}(b) and~\ref{rings_up}(b), that the inflection of the phasefront curvature changes according to the focal length of the lens used (\texttt{$\pm\,500\,$}mm) respectively. 
Within the LTP interferometry research and development, this instrument will be used during manufacture of the optical bench interferometer flight model. The lenses of the fiber injectors can be adjusted such that the difference in curvature between the two interfering wavefronts are minimized by using the real-time phasefront read out provided by this phasemeter. A phasefront measurement was already conducted at the optical bench engineering model for LTP\,\cite{ghh1,ghh2}. These results are presented in Figure~\ref{ob-phase} and clearly show an inhomogeneous phasefront, which can be attributed to non-optimal adjustment of the lenses in the two fiber injectors. A further test was done on a table-top interferometer in order to adjust the lenses of two commercial fiber injectors by using this instrument. The aim of this adjustment was to match the parameters of the interfering beams and to obtain a homogeneous flat phasefront. The result of this experiment is shown in Figure~\ref{flat}.
A considerable improvement was achieved in matching the shape of the two beams (compare the phase scale of the plots in Figures~\ref{ob-phase} and~\ref{flat}).
It can be seen that over a surface of approximately 1.2\,mm\texttt{$\times$}1.2\,mm the phasefront shows a considerably homogeneous spatial profile. We analyzed a circular section of approximately 1\,mm diameter at the beam center. The standard deviation of these data is 3.49\,mrad, which is very close to the measured sensitivity of the instrument. This value corresponds to a spatial resolution of 590\,pm, and amounts to a considerable improvement in the correction of the beam shapes.
\section{Conclusions}
\label{conclu}
We have developed an instrument which allows real-time phasefront detection and mode-mismatch characterization of two interfering beams in a heterodyne interferometer. A rms noise level of 2.96\,mrad, which corresponds to a wavefront roughness of 500\,pm, was obtained experimentally. This makes it possible to optimize the beam shapes by adjusting the optical components in real-time with the help of the data displayed onto the graphical user interface (5 to 6 data displays per second). The results shown in Figures~\ref{rings_down} to~\ref{cylindric} demonstrate the proper functionality of the instrument. By using well matched wavefronts, this method can also be used to analyze, and measure accurately, surfaces and optical components down to subnanometer levels. It is planned that this instrument be used in the manufacturing of the flight model of the optical bench for the LISA Technology Package.
\section*{Acknowledgements}
\label{thanks}
The authors kindly thank Albrecht R\"udiger, Benjamin Sheard, Paul Cochrane, Alexander Bunkowski, James DiGuglielmo, and Martin Hewitson for their valuable contributions in reviewing and improving various parts of the manuscript.
%
%%%%%%%%%%%%%%%%%%%%%%% References %%%%%%%%%%%%%%%%%%%%%%%%%
%

%
\end{document}